\documentclass[12pt]{article}
\usepackage{amsmath}
\usepackage{amsfonts}
\usepackage{epsfig}
\newcommand{\beq}{\begin{equation}}
\newcommand{\eeq}{\end{equation}}

\newcommand{\edo}{\end{document}}

\newcommand{\ty}{\tilde{y}}

\title{\bf Entropy from inclusive scattering}
\author{M.A.Braun  \\
S.Peterburg State University, Russia}

\begin{document}
\maketitle

\abstract
{Introduction of probabilities from multiple inclusive cross-sections is studied
in  unitary pomeron exchange models with no interaction between the pomerons themselves.
Discrepancy is observed between the emerging inelastic cross-section and the one following from unitarity.
It diminishes with energy and disappears in the high-energy limit.
Probabilities and entropy are calculated numerically.
 As a function of energy squared  $s$
the entropy grows as $\ln\ln s$ rather than as $\ln s$,
contrary to naive predictions. For the BFKL pomeron the entropy
abruptly changes its slope in the rapidity region where consistency
of the model is approximately restored.}

\section{Introduction}
Considerable attention has recently been devoted to the
probabilistic interpretation of the scattering cross-sections
and the following entropy. At first sight such an approach cannot
meet with any difficulties. In fact the quantum mechanics and scattering
theory are based on probabilities for the scattering to occur and its
cross-sections. The latter are nothing but probabilities to observe
scattering products in some volume of the overall phase space. The problem
arises when one wants to construct the entropy, standardly determined
from probabilities related to discrete elementary events. The cross-sections
however  are rather probability densities in the continuous characteristics
of the scattering process. So wishing to construct the entropy one has to
somehow discretize the scattering characteristics,

This can be attempted in different ways. Perhaps the simplest is to
split the total phase volume $\Omega$, which includes number of particles,
momenta, spins, charges etc, into a discrete number of cells $\Delta\Omega$.
Transitions to such cells will form a discrete set of events with
well defined probabilities from which the entropy could be directly
constructed. However such probabilities and entropy will of course strongly
depend on the cell volume $\Delta\Omega$. Once in the desire to eliminate
this dependence one tries to make the limit $\Delta\Omega\to 0$ both
the probability and entropy will acquire infinite terms or factors.

This approach was taken in a series of papers ~\cite{1}-\cite{5}.
In particular in ~\cite{3,5} to somewhat alleviate the problem the
authors considered elastic scattering to states with given orbital momenta
thus forming a partially discrete set of final states. However passing to
orbital momenta does not eliminate continuity of the spectra
in its totality.
As a result the resulting probabilities depend on one infinite factor,
which the authors try to somehow regularize.

A more promising direction taken in most studies is to discretize the
scattering data using the natural discrete parameter, the number
of produced particles $n$.
The total cross-section can be presented as a sum of contributions
coming from $n$ emitted particles
\beq
\sigma^{tot}=\sum_{n=0}\sigma_n\label{sign}\eeq
$\sigma_0=\sigma^{el}$ corresponding to absence
 of emitted particles,
The sum of all terms with $n\geq 1$ gives the total inelastic cross-section
\[\sum_{n=1}\sigma_n=\sigma^{in}.\]
One can define the probabilities to find exactly $n$ emitted
particles by
\[P(n)=\frac{\sigma_n}{\sigma^{tot}},\ \ \sum_{n=0}P(n)=1,\ \
P(n)\geq 0.\]
Once $\sigma_n$ are known probabilities and entropy
can be directly constructed.

Note that there is an alternative way to construct probability from
the scattering data. One can start from integrated
$k$-fold inclusive cross=sections and consider them as probability momenta:
\[\nu(k)= \int (dq'_1)...(dq'_k)\frac{I_k(q'_1,..q'_k)}{\sigma^{tot}}=
\sum_n \frac{n!}{(n-k)!}\frac{\sigma_n}{\sigma^{tot}}=
\sum_{n\geq k}\frac{n!}{(n-k)!}P(n).\]
Here
\[(dq)=\frac{dyd^2q_\perp}{8\pi^3}.\]
Note that with this definition
\[\nu(0)=\sum_{n=0} P(n)=1\]
The inverse relation allows to find $P(n)$ once the momenta $\nu(k)$
are known:
\beq
P(n)=\frac{1}{n!}\sum_{k\geq n}(-1)^{k-n}\frac{\nu(k)}{(k-n)!}.
\label{pvianu}\eeq
It is important that both  ways of constructing $P(n)$, from $\sigma_n$
and from $\nu(k)$ should give  the same probabilities. This implies
that the set of $k$-fold inclusive cross=sections should agree
with the set of partial cross-sections $\sigma_n$. As we shall discover
in the following
this is in fact a strong restriction on the consistency of
proposed scattering models.

Construction of probabilities from the partial cross-sections $\sigma_n$
has long been considered in the models based on the QCD pomeron exchange.
The probability of a state with $n$ produced particles was customary associated
with a convolution of the probability to find a given number of
active pomerons ("cut pomerons") with the probability to find
a given number of particles produced from the pomeron.

In  the QCD constructive representation of the scattering amplitudes
in terms of pomeron exchanges is hardly possible generally.
So the authors restricted
themselves to simplified models, to fan diagrams contribution or
"big loops", which are actually the combination of fans from the projectile and from
the target.  Significant results were achieved in
numerous studies devoted to
the one dimensional models with no transverse space ~\cite{6}-\cite{11} starting from the
seminal paper ~\cite{6}. It was found that the probabilities could be naturally
constructed and found
to satisfy certain evolution equation. Its  solution gives the entropy. which
rises logarithmically as its logaritym.

In the real 2-dimensional transverse
world in
~\cite{8},\cite{12}-\cite{16} the set of fan diagrams generated ny the  Baltitski-Kovchegov
equation ~\cite{17,18} was studied to find the distribution of cut pomerons.
For this distribution an appropriate evolution equation was established
~\cite{12}. Due to its complexity various approximate schemes of its solutions were
proposed with the final results for probabilities and entropy in ~\cite{15,16}.
The found entropy was growing with energy possibly in the linear way.

Two questions arise with these pomeron studies.
First due to the simplified scattering picture actually corresponding to
DIS the mentioned
problem of consistency did not arise.
Interaction with the projectile was taken to be weak, so that
the amplitudes were not constrained by the
unitarity relation.

Second, transition from pomerons to produced particles (gluons)
in the QCD cannot be limited to the convolution with the distribution inside pomerons only.
As was shown in ~\cite{19} apart from production from within the pomeron
an extra contribution takes place, which can be interpreted as emission
from the vertex. We do not know of any studies of this contribution
beyond the single and double gluon production in ~\cite{19,20}.

This motivated us to study an essentially much simpler
model
with no interaction between pomerons, corresponding to scattering diagrams
shown
in Fig. 1. with either BFKL pomeron or local Regge-Gribov (RG) pomeron
exchanges. Considerably simpler than fan diagrams, the amplitude generated by
sequential
pomeron exchanges is unitary and so presents an ideal laboratory to
study the relation between the cross-section $\sigma_n$ and $n$-fold
inclusive ones to see the consistency of the model. As we will show
emission probabilities from a pomeron are not consistent with
the total inelastic cross-section at finite energies. This inconsistency
was first noted long ago ~\cite{21}. The emission
cross-section  results somewhat
smaller. However  consistency is fully restored in the limit
of infinitely high energies, which is not so astonishing
since the pomeron theories are oriented to this limit.
 The magnitude of  the found inconsistency diminishes
exponentially with energy, the rate depending on the production
intensity. As we found, for the BFKL pomeron exchanges the inconsistency
 goes down to around
2\% at rapidity $y=20$ but is substantially larger at smaller energies.
\begin{figure}
\begin{center}
\epsfig{file=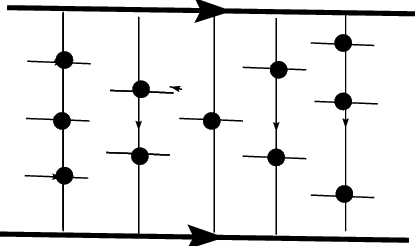, width=8 cm}
\caption{Multiple pomeron exchanges}
\end{center}
\label{fig1}
\end{figure}

\section{Regge-Gribov (RG) pomeron}
\subsection{Probabilities}

In this section we consider the most simple case of multiple
exchanges of the RG pomeron found as a pole
in the complex orbital momentum plane.
Its multiple exchanges generate cuts in this plane,
This model was exhaustively discussed long ago, see e.g. ~\cite{22}-\cite{30}.
It was also generalized there to include excited participant hadron states
and different forms of particle production to better describe the data.
For our purpose these generalizations are irrelevant and we will consider the
model in its simplest version corresponding to Fig. 1.
We will reproduce some basic formulas to fix our notations.

Writing the pomeron trajectory in the customary way
\[\alpha(t)=1+\Delta+\alpha't\]
we find the elastic scattering amplitude given by the single pomeron exchange as
\beq
{\cal A}^{(1)}(s,t)=2isRe^{-\lambda t}\label{a1}.\eeq
Here $R$ does not depend on $t$
\[ R=\frac{3\pi g^2}{4s_0}\Big(\frac{s}{s_0}\Big)^\Delta\]
and includes coupling to the participants $g^2$ (dimensionless)
and energy scale $s_0$. The total slope $\lambda$
\[\lambda(s)=\lambda_0+\alpha'\ln \frac{s}{s_0}\]
includes contribution $\lambda_0$ from the impact factors.

To pass to the eikonal approximation one rewrites (\ref{a1}) as the
integral in the impact parameter space
\[{\cal A}^{(1)}(s,t)=2 is\int d^2be^{i{\bf rb}}\rho(b)\]
where
\[\rho(b)=re^{-b^2/4\lambda},\ \ r=\frac{R}{4\pi\lambda}.\]
The eikonal approximation to the total amplitude with any number of
pomeron exchanges is then
\[{\cal A}(s,t)=2 is\int d^2be^{{\bf rb}}
\Big(1-e^{\rho(b)}\Big).\]
Taking $t=0$ we get the total cross-section
\[\sigma^{tot}(s)=2 \int d2^b
\Big(1-e^{-\rho(b)}\Big).\]
Transforming to integration variable $\rho$ we find
\beq
\sigma^{tot}=8\pi\lambda F(r),\label{sigma}\eeq
where
\beq
F(r)=\int_0^r\frac{d\rho}{\rho}\Big(1-e^{-\rho}\Big)=
-\sum_{n=1}\frac{(-r)^n}{nn!}=\ln r+{\bf C}-{\rm Ei}(-r).\label{fr}\eeq
Here Ei is the exponential integral function and ${\bf C}$
is Euler's constant.
Separate terms in the sum over $n$ correspond to cross-sections
with $n$ pomeron exchanges
\beq
\sigma^{(n)}=-8\pi\lambda \frac{(-r)^n}{nn!}\label{sigpom}\eeq
(not to confuse with $\sigma_n$ in (\ref{sign}), which refers
to $n$ produced particles).

Using the AGK rules ~\cite{agk} one can find the cross-section from
$n$ exchanged pomerons with $k\leq n$ of them cut
\beq \sigma^{(nk)}=8\pi\lambda\frac{1}{2}\frac{(2r)^n}{nn!}C_n^k(-1)^{n-k},\label{snk}\eeq
since each cut pomeron contributes $2r$ and an uncut one (-2r).
Note twice smaller coefficient in front reflecting the fact that
$\sigma^{(nk)}$ comes from twice the imaginary part of the amplitude.
Summing $\sigma^{(nk)}$ over all $k\geq 1$ one obtains the
inelastic part
\[\sigma^{(in,n)}=-8\pi\lambda\frac{1}{2}\frac{(-2r)^n}{nn!}\]
and summing over $n$ obtains the total inelastic cross-section
\beq\sigma^{(in)}=
8\pi\lambda \frac{1}{2}F(2r).\label{sigin}\eeq
The  elastic cross-section is evidently
$\sigma^{el}=\sigma^{tot}=\sigma^{in}$.

If one sums $\sigma^{(nk)}$ over $n$ then one obtains the cross-section
from exactly $k$ cut pomerons
\[\sigma_{cut}^{(k)}=8\pi\lambda\frac{1}{2}\,\frac{1}{k!}
\sum_{n=k}^\infty\frac{(2r)^n}{n(n-k)!}(-1)^{n-k}=
8\pi\lambda\frac{1}{2}\,\frac{(-2r)^k}{k!}\sum_{n=0}^\infty
\frac{(-2r)^n}{(n+k)n!}.\]
Of course
\[\sum_{k=1}^\infty\sigma^{(k)}=\sigma^{(in)}.\]
So  interchanging indices $n$ and $k$ we can introduce the probability
to have exactly $n$ cut pomerons
\[P_{pom}(n)=\frac{\sigma_{cut}^{(n)}}{\sigma^{(in)}}=
\frac{1}{F(2r)}\frac{(-2r)^n}{n!}
\sum_{k=0}^\infty\frac{(-2r)^k}{(n+k)n!}.\]
One can express the sum
\[
G_n(-x)=\sum_{k=0}^{\infty}\frac{(-x)^k}{(k+n)k!}\]
via the incomplete Gamma-function.
Presenting
\[\frac{1}{k+n}=\int_0^\infty dz e^{-z(k+n)}\]
and passing to integration variable
\[y=xe^{-z}, dz=-\frac{dy}{y}\]
we get
\beq
G_n(-x)=x^{-n}\int_0^xdyy^{-1+n}e^{-y}=
x^{-n}\Gamma(n,x),\label{gnx}\eeq
where $\Gamma(n.x)$ is the incomplete $\Gamma$-function.
So we find the cut pomeron distribution as
\beq
P_{pom}(n)=\frac{1}{n!}\frac{\Gamma(n,2r)}{F(2r)}, \ \ n\geq 1
\label{pomn1}\eeq
with
\[\sum_{n=1}P_{pom}(n)=1.\]
Note that this is only the distribution of cut pomerons. To find
the distribution in produced particles one expects to convolute
the distribution of pomerons with the distribution of particles produced
from a single pomeron. Presently we prove that this expectation is
justified.

We turn to the determination of the probability to find $n$  produced
particles via $k$-fold inclusive cross-sections as discussed in
Section 1.
Due to the AGK cancelations for the $k$-fold
inclusive cross-section it is sufficient to consider diagrams with exactly
$k$  exchanges with particle emitted from each pomeron, Fig. 1.

Consider emissions from a single pomeron.
Let the vertex for the emission of a particle at rapidity $y$ and with
transverse momentum $q_\perp$ be $f(q)$. Assuming that emissions  are independent,
for $k$ emissions
we will find the pomeron propagator multiplied by
\[ \Big(\int \frac{d^2q}{(2\pi)^2}q^2f(q)\Big)^k=c^k\]
Further integration over rapidity $y$ will give the total emission
factor $c^kY^k/k!=j^k/k!$. The corresponding probability momenta
$\nu(k)$ will be obtained after all permutations of produced particles,
so that $\nu(k)=j^k$. Using (\ref{pvianu}) we then immediately find that the probability
to find $n$ particles in a single pomeron is the Poisson distribution
\beq
P_{part/pom}(n)=e^{-j}\frac{j^n}{n!},\ \ j=cY.\label{poisson}\eeq
Factor $c$ evidently
has the meaning of multiplicity per unit  rapidity from
the single pomeron exchange.

Now
take a diagram with $k$ exchanged pomerons producing $n\geq k$
particles, Fig. 1. We assume that $j$-th pomeron  emits $n_j$ particles
so that
\beq
\sum_{j=1}^k n_j=n,\label{sumj}\eeq
The contribution to the $n$-fold inclusive cross-section will be
\[I_{nk}=-\frac{1}{2}j^n(-2)^k\sigma^{(k)}\sum_{n_1,...n_k}
\frac{n!}{n_1!...n_k!},\ \ \sum_{j=1}^k n_j=n,\ \ n\geq k\geq 1.\]
The factorials in the denominator reflect the fact that integration
over rapidities of $n_j$ particles emitted from the $j$-th pomeron
gives factor $Y^{n_j}/n!$. Note that taking $n=1$
we find the multiplicity
\beq
\mu=j\frac{\sigma^{(1)}}{\sigma^{tot}}.\label{mu}\eeq

To do the sum over all $n_j$ we introduce the  Kroneker symbol
\[\delta_{n0}=\frac{1}{2\pi i}\int_C dzz^{n-1}\]
where integration goes over the circle around zero.
So the sum over $n_j$ under condition (\ref{sumj}) can be written as
\beq
\frac{1}{2\pi i}\sum_{n_1,...n_k}
\frac{n!}{n_1!...n_k!}\int_C dzz^{\sum_{j=1}^k n_j-n-1}\label{sumj1}
\eeq
where one can sum over each $n_j$ from unity to infinity.
Each sum gives
\[\sum_{n_j=1}^\infty \frac{z^{n_j}}{n_j!}=e^z-1\]
and the sum (\ref{sumj1}) transforms into the integral
\[ \frac{1}{2\pi i}\int_C dzz^{-n-1}\Big(e^z-1\Big)^k=
\frac{1}{n!}\Big(\frac{d}{dz}\Big)^n\Big(e^z-1\Big)^k\Big|_{z=0}
=\frac{1}{n!}\sum_{m=0}^k(-1)^{k-m}C_k^mm^n.\]

We obtain the probability moments
\[ \nu(k)=aj^k\sum_{l=1}^k
\frac{(2r)^l}{ll!}
\sum_{m=1}^l(-1)^{l-m}C_l^mm^k,\ \ k\geq 1\]
where
\[a=\frac{4\pi\lambda}{\sigma^{tot}}=\frac{1}{2F(r)}\]
and  $F(r)$ is given by (\ref{fr}).
To transform the $k$-dependence into  pure power-like
we introduce the appropriate $\theta$-function and allow
to sum over $l$ up to infinity: for $k\geq 1$
\[ \nu(k)=aj^k\sum_{l=1}^\infty\theta(k-l)
\frac{(2r)^l}{ll!}
\sum_{m=1}^l(-1)^{l-m}C_l^mm^k\]
\[=a\frac{1}{2\pi i}\int_{-\infty}^\infty\frac{d\xi}{\xi-i0}
e^{i\xi(k-l)}
\sum_{l=1}^\infty\frac{(2r)^l}{ll!}
\sum_{m=1}^l(-1)^{l-m}C_l^m(jm)^k.\]

Now we calculate the probabilities $P(n)$ by Eq. (\ref{pvianu})
\[P(n)=\frac{1}{n!}\sum_{k=n}(-1)^{k-n}\frac{\nu(k)}{(k-n)!}\
\ n\geq 1.\]
This will change the $k$ depending factor
\[\Big(jme^{i\xi}\Big)^k\to
\Big(jme^{i\xi}\Big)^n\exp\Big(-jme^{i\xi}\Big).\]

So we get for $n\geq 1$
\[P(n)=
\frac{a}{n!}\frac{1}{2\pi i}\int_{-\infty}^\infty\frac{d\xi}{\xi-i0}
\sum_{l=1}^\infty\frac{(2r)^l}{ll!}e^{-i\xi l}
\sum_{m=1}^l(-1)^{l-m}C_l^m
\Big(jme^{i\xi}\Big)^ne^{-jme^{i\xi}}.\]
Changing the order of $l$ and $m$ summations gives
\[P(n)=
\frac{a}{n!}\frac{1}{2\pi i}\int_{-\infty}^\infty\frac{d\xi}{\xi-i0}
\sum_{m=1}^\infty\frac{1}{m!}
\Big(cYme^{i\xi}\Big)^ne^{-cYme^{i\xi}}X,\]
where
\[X=\sum_{l=m}^\infty(-1)^{l-m}
\frac{(2re^{-i\xi})^l}{l(l-m)!}.
\]
We do the sum over $l$ denoting
\[z=2re^{-i\xi}\]
and putting $l=m+l'$. We get
\[
X=z^m\sum_{l'=0}^\infty(-1)^{l'}\frac{z^{l'}}{(l'+m)l'!}=
\Gamma(m,z),\]
where $\Gamma(m,z)$ is the incomplete $\Gamma$-function.
So we find
\[P(n)=
\frac{a}{n!}\frac{1}{2\pi i}\int_{-\infty}^\infty\frac{d\xi}{\xi-i0}
\sum_{m=1}^\infty\frac{1}{m!}
\Big(jme^{i\xi}\Big)^ne^{-jme^{i\xi}}
\Gamma\Big(m,2re^{-i\xi}\Big).\]

Now our desire is to do the integral over $\xi$.
To this end we have to study analytic properties of the function
of $\xi$
\[ G(\xi)=\Big(jme^{i\xi}\Big)^ne^{-jme^{i\xi}}
\Gamma\Big(m,2re^{-i\xi}\Big).\]
Obviously it is analytic in the whole complex $\xi$ plane.
The $\Gamma$ function is
\[
\Gamma\Big(m,2re^{-i\xi}\Big)=\int_0^{2re^{-i\xi}}dtt^{m-1}e^{-t}=
\Big(2re^{-i\xi}\Big)^m\int_0^1dt t^{m-1}e^{-2tre^{-i\xi}}\]
Take $\xi=\xi_1+i\xi_2$ with real $\xi_{1,2}$.
Then under the integral over $t$ the behavior at large $\xi_2$
is determined by double exponential factors
\[\exp(-jme^{i\xi_1-\xi_2})\exp(-te^{-i\xi_1+\xi_2})\]
If $\xi_2\to\infty$ then the first is finite and the second very
strongly goes to zero making the total integrand vanish.
If $\xi_2\to -\infty$ then vice versa the first factor very
strongly goes to zero and the second is finite
so that the total integrand again vanishes.
At first sight $G(\xi)$ goes to zero in both directions
upwards and downwards in the $\xi$-plane. However as $\xi_2\to -\infty$
the exponent  in the integrand of $\Gamma $-function vanishes
and the integral diverges. So in fact one can only shift the
contour upwards and  has to take residue at $\xi=0$.
Thus we find finally
\beq
P(n)=
\frac{a}{n!}\sum_{m=1}^\infty\frac{1}{m!}
(jm)^ne^{-jm}\Gamma(m,2r),\ \ n\geq 1\label{pnf}\eeq
and as before
\[P(0)=1-\sum_{n=1}P(n)\]
One can check that $<n>=\sum nP(n)$ gives (\ref{mu}).
Probability $P(n)$ given by (\ref{pnf}) has the expected meaning. namely if one forgets that in
it both $n$ and $m$ are greater than zero  then it is just a convolution of
the probability to find $m$ cut pomerons with the Poisson probability to
find $n$ particles from each pomeron:
\[P=P^{*poisson}_{part/pom}\otimes P^*_{pom}\]
where asterisks mean dropping terms with no scattering and no emission.
This is a particular realization of the two-stage scenario of particle
emission ~\cite{bvp}. Thus we have rigorously proved the customary
transition from the probability for cut pomerons to that for
$n$ particle production.

Using (\ref{pnf}) and summing over $n\geq 1$ we can express
$\sigma^{in}/\sigma^{tot}$ via standard functions.
\[
\frac{\sigma^{in}}{\sigma^{tot}}=\sum_{n=1}^\infty P(n)=
a\sum_{m=1}^\infty\frac{1}{m!}e^{-jm}\Gamma(m,2r)
\sum_{n=1}^\infty\frac{1}{n!}(jm)^n=
a\sum_{m=1}^\infty\frac{1}{m!}e^{-jm}\Gamma(m,2r)
\Big(e^{jm}-1\Big)\]\beq
=\frac{F\Big(2r(1-e^{-j})\Big)}{2F(r)}\label{ratio}\eeq

Rigorous consistence of the model requires the numerator to be
equal to
$F(2r)$, which implies $j=cY\to\infty$, that is the number of particles
emitted from each pomeron should be infinitely large.
This implies that consistency is achieved only at infinitely high
energies. This is not astonishing, since pomeron models are supposed
to be valid only in the high  energy limit. However violations of
consistency vanish very fast with energy, as $\exp(-cY)$.
With $c$ of the order unity, as extracted from the data, already at
$y=10$ this violation is $\sim 10^{-4}$.

One can try to estimate the probabilities $P(n)$ for large $j$.
The asymptotic of the incomplete $\Gamma$-function can be estimated by
the stationary point method to be roughly
\[\Gamma(n,x)=\theta(x-n)\Gamma(n).\]
Of course is a very crude estimate: in fact  $\Gamma(n,x)$ is much less
at $n>x$ but not zero.
Using this we have
\[
P(n)=
\frac{a}{n!}\sum_{m=1}^{2r}\frac{1}{m}
(jm)^ne^{-jm}.\label
{pnfin}\]
If $n>1$ one can start summation from $m=0$. One can  approximate the sum by
the integral over $m$
\[
P(n)=
\frac{a}{n!}\int_0^{2r}\frac{dm}{m}
(jm)^ne^{-jm}=
\frac{a}{n!}\int_0^{2jr}\frac{dx}{x}x^ne^{-x}.\]
As $j>>1$
\beq
P(n)\simeq \theta(r-n)\frac{1}{n}\,\frac{1}{2F(r)}\simeq
\theta(2r-n)\frac{1}{n}\,\frac{1}{2\ln r}.
\label{aspn2}\eeq
Restriction $n<2r$ comes from summation over $n$, which should give 1/2.
This asymptotic does not depend on $j$. Its $Y$-dependence comes
only from normalization and number $n\sim r$ starting from which
the probability abruptly goes down.
However  our numerical studies  in the next subsection show that this  asymptotical
estimate is too crude and inadequate at large $n$,

With the found probabilities  the
(von-Neumann) entropy can be found in the standard manner
\beq
S(Y)=-\sum_{n=0}P(n)\ln P(n).\label{ent1}\eeq
From our crude asymptotic (\ref{aspn2}) one can conclude at large energies the entropy
becomes a linear function
\beq
S(Y)\sim\beta Y\label{asym}\eeq
Indeed
with the asymptotic (\ref{aspn2}) we have
\[S(y)=-\sum_n P(n)\ln P(n)=
-\frac{1}{2\ln r}\sum_{n=1}^{2r}\frac{1}{n}\ln\frac{1}{n}\simeq
\frac{1}{2\ln r}\int_1^{2r}\frac{dn}{n}\ln n\simeq
\frac{1}{4}\ln r
\simeq \frac{1}{4}\Delta Y.\]
So one finds  a linear asymptotic (\ref{asym}) with coefficient
$\beta=\Delta/4$.
Note that the linear growth with $Y$ is also found in
in zero-dimensional models ~\cite{7,8} in which however $\beta$
seems rather the
intercept of  the hard pomeron significantly greater than ours.
As  we shall presently see however this behavior
is not seen in our  calculations in the interval $0< Y<30$
due to very poor approximations used in its derivation.
Actually we will find that at high energies the entropy calculated
numerically from
(\ref{pnf}) rises as $\beta\ln Y$, much slower than (\ref{asym})

%%%%%%%%%%%%%%%%%%%%%%%%%%%%%%%%%%%%%%%%%%%%%%%%%%%%%%%%%%%
%%%%%%%%%%%%%%%%%%%%%%%%%%%%%%%%%%%%%%%%%%%%%%%%%%%%%%%%%%%%%%%%
\subsection{Numerical studies}
We choose the more or less standard intercept and slope of the
pomeron
\beq
\alpha_p(0)-1=\Delta=0.13,\ \ \alpha'=0.2\ GeV^{-2}.\label{par}\eeq
For the impact parameter slope we take 2.0 $GeV^{-2}$
(corresponding to the inverse mass of the $\rho$ meson).
As to coupling $g$ we take it to approximately agree with the
proton-proton total cross-section.
$g^2=6$.
Cross-sections generated in our model for $Y\leq 30$ are shown
in Fig. 2.
\begin{figure}
\begin{center}
\epsfig{file=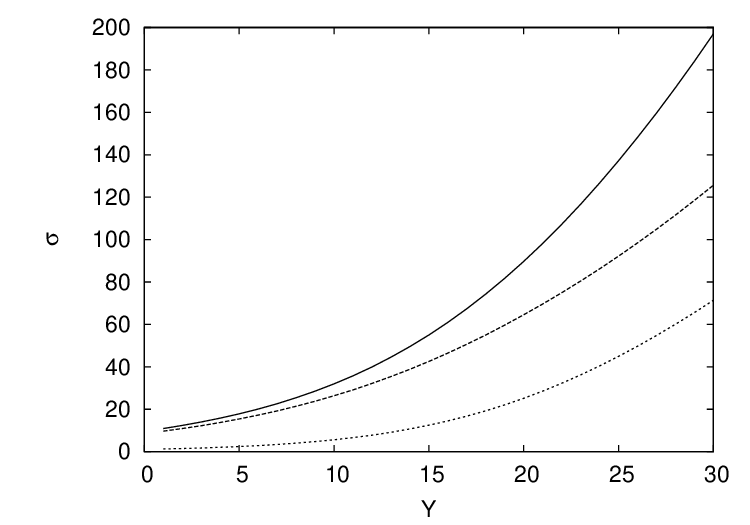, width=8 cm}
\caption{Total (upper curve), inelastic (middle curve)
and elastic (lower curve) cross-sections at different rapidities}
\end{center}
\label{fig2}
\end{figure}
The value of the characteristic dimensionless rescattering parameter
$r(Y)$ is demonstrated in Fig. 3.
\begin{figure}
\begin{center}
\epsfig{file=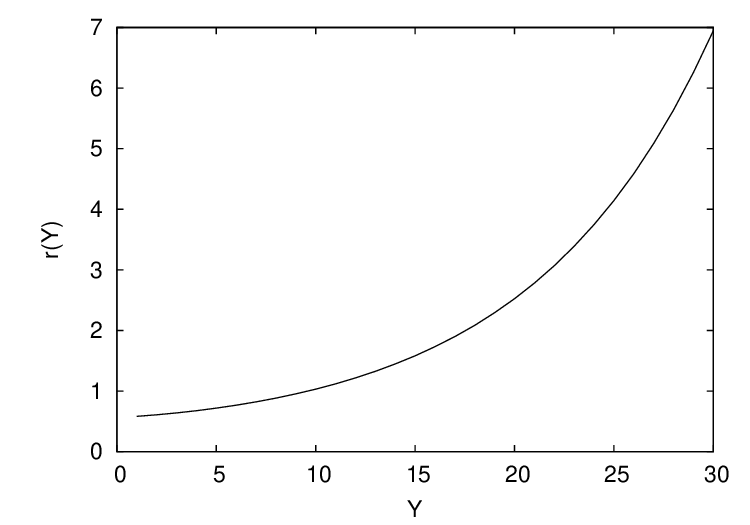, width=8 cm}
\caption{ Rescattering parameter $r(Y)$}
\end{center}
\label{fig3}
\end{figure}
As we see one cannot say $r>>1$ in our energy interval.

With  these parameters  we calculated probabilities
$P(n)$  with $j=cY$ and $c=1$ (multiplicity per unit rapidity).
In Eq.(\ref{pnf}) we summed over
$n,m\leq n_{max}$ with different $n_{max}$ from 40 to 3200.
It turned out that at $y\leq 30$ the probabilities $P(n)$
for $n\leq 40$ are
well described already with $n_{max}=40$. However the entropy is
stabilized only at $n_{max}\geq 800$.

To exclude influence of normalization
in Fig. 4 we show probabilities $P(n)$
multiplied by $\sigma^{tot}$.
The curves with maxima at higher $Y$ correspond to  $Y=10,20$ and 30
respectively. Their behavior is quite different from our
asymptotical estimates. Our curves have maxima which are shifted
to higher $n$ with the growth of energy with values of these
maxima slowly diminishing.
\begin{figure}
\begin{center}
\epsfig{file=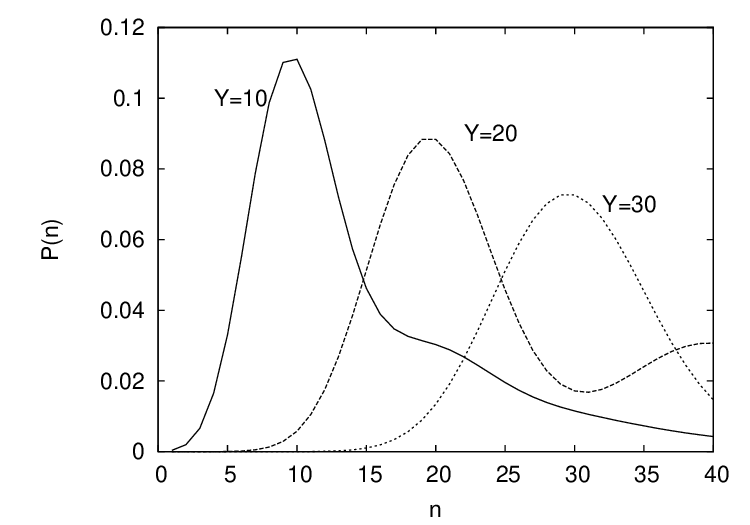, width=8 cm}
\caption{Probabilities $P(n)$
multiplied by $\sigma^{tot}$ for $Y=10,20$ and 30.}
\end{center}
\label{fig4}
\end{figure}

The entropy  is shown in Fig.5. As one observes
its behavior with energy is well described by  $\ln\ln s$
dependence:
\beq
S(Y)=\beta\ln Y, \ \beta=0.85\label{ey}.\eeq.
It is softer than the behavior following from our crude
estimation (\ref{asym}) or found in one-dimensional models.
\begin{figure}
\begin{center}
\epsfig{file=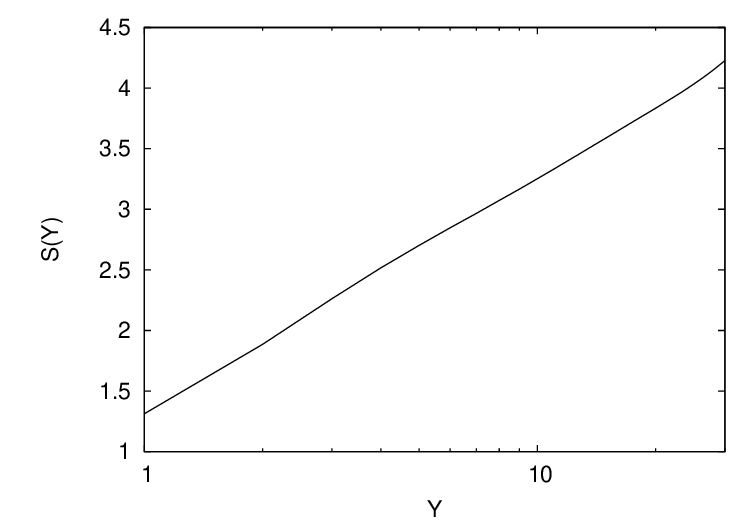, width=8 cm}
\caption{Entropy $S(Y)$.}
\end{center}
\label{fig5}
\end{figure}

%\edo

\newpage
\section{ BFKL pomeron}
\subsection{Cross-sections and cut pomerons}
Our second model is the  BFKL pomeron exchanges corresponding to the same
Fig. 1 but with BFKL pomeron rungs.
%%%%%%%%%%%%%%%%%%%%%%%%%%%%%%%%%%%%%%%%%%%%%%%%%%%%%%%%%%
%%%%%%%%%%%%%%%%%%%%%%%%%%%%%%%%%%%%%%%%%%%%%%%%%%%%%%%%%%%
It is straightforward
to generalize our  treatment of the previous section
to BFKL pomerons.
However in doing so one should have in mind two important properties of the
BFKL pomeron which differ it from the
  RG one. First, the BFKL pomeron has no slope:
its intercept $\alpha(t)=1+\Delta$ does not depend on $t$.
 The $t$ dependence is then provided exclusively by the impact factors.
However it is independent of energy in contrast to  the RG pomeron
and so it is somewhat external to the model. Second, the emission from the
BFKL pomeron is fully determined  as studied in ~\cite{bt} long ago,
whereas in the RG model it is rather external to the model to be
adjusted to comply with the cross-sections.

Having in mind these two features of the BFKL pomeron we
can eikonalize the scattering amplitude in the same way as
previously in  Section 2. for  the RG pomeron.

Recalling the cross-section
 coming from the single BFKL pomeron exchange  ~\cite{bt} we have the
 corresponding amplitude
\[2{\rm Im}{\cal A}(s,t)=2s\frac{32}{9}\alpha_s^2R_1R_2 e^{\Delta Y}
\sqrt{\frac{\pi}{aY}}e^{-\lambda k^2}.\]
where $t=-k^2$ and $\lambda$ comes from the impact factors and does not depend on
$Y$.
Dimensionless $\Delta$ and $a$ are expressed by $\alpha_s$ and
given by
\beq
 \Delta=(12\alpha_{s}/\pi)\ln 2,\ \
a=(42\alpha_{s}/\pi)\zeta(3).
 \label{dela}\eeq
Comparing with the similar amplitude for the RG pomeron
(\ref{a1}) we find the  scattering factor $r(Y)$ for the
BFKL pomeron
\beq
r=\frac{32}{9}\frac{\alpha_s^2R_1R_2e^{\Delta y}}
{4\pi\lambda} \sqrt{\frac{\pi}{aY}}.\label{rbfrl}\eeq

Once $r$ is determined one can repeat the eikonalization procedure
for the BFKL pomeron exactly in the same manner as for the RG pomeron.
One obtains then the same formulas
for the cross-sections (\ref{sigma}) and (\ref{sigin})
and cut pomeron probabilities (\ref{pomn1})
with the BFKl expressions for
$r$ and $\lambda$.

\subsection{Gluon emission from the single BFKL pomeron}
To find the final probabilities $P(n)$ for $n$-gluon production we
have to find them from the single cut BFKL pomeron
 illustrated in Fig. 6.
\begin{figure}
\begin{center}
\epsfig{file=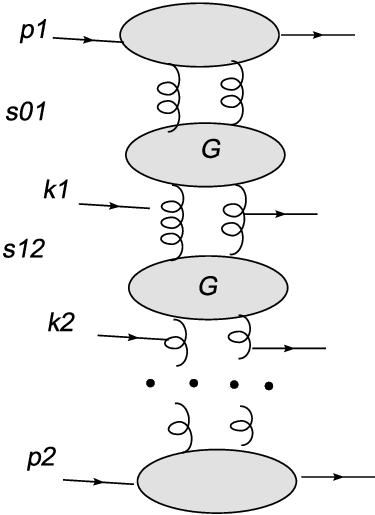, width=6 cm}
\caption{Gluon production  from the BFKL chain}
\end{center}
\label{fig6}
\end{figure}
This was studied in  ~\cite{bt}
from which our basic equations can be borrowed. Using the explicit
expressions for the BFKL forward Green function and vertex for emission
of a gluon from it one finds the  inclusive cross-section
for  production of $n$ gluons ar rapidities and transverse momenta
$(y_i,k_i)$, $i=1,2,...n$, where $Y/2>y_1>y_2>,,,y_n>-Y/2$ with all
differences $y_{i,i+1}=y_i-y_{i+1}$  positive and large, in  the form
  \beq
I_{n}(y_i,k_i)=
\frac{64}{9}\alpha_{s}^{2}R_{1}R_{2}
\prod_{i=1}^{n}\frac{3\alpha_{s}}{4\pi^{3}
k_{i}^{2}}(s\prod_{i=1}^{n}k_{i}^{2})^{\Delta}
(\sqrt{\pi/a})^{n+1}\prod_{i=0}^{n}y_{i,i+1}^{-1/2}\,J_{n},\label{inbt}
  \eeq
  where
  \beq J_{n}=
\prod_{i=1}^{n}\int_{\zeta_{i}}^{\infty}d\alpha_{i}
\exp
\left(-\sum_{j=0}^{n}\frac{(\alpha_{j}-\alpha_{j+1})^{2}}
{4ay_{j,j+1}}\right).
 \eeq
Here
  $
  \zeta_{i}=\ln k_{i}^{2}
  $
  and it is assumed that
$ \alpha_{0}=\alpha_{n+1}=0 $.
Note the appearance of factors $k_{i}^{2\Delta}$ in this direct derivation ~\cite{bt}..
They make cross-section  coming from (\ref{inbt}) integrable at small $k_{i}^{2}$  and thus
infrared stable. However  retaining these
factors actually leads beyond the accuracy
 adopted in the lowest order (LO) hard pomeron model, in which terms
 containing powers of
$ \alpha_{s}\ln k^{2} $ are systematically neglected.
Still the LO BFKL model in any case is not applicable at small $k_\perp$ and needs some infrared
cutoff. The direct derivation followed in ~\cite {bt} provides a "natural" cutoff $\Delta$.
However the leading order BFKL model describes emission of massless gluons and neglects energy conservation.
So to bring the model closer to physically meaningful predictions the cutoff can be changed to a more reasonable 
penomenological one $\Delta_{phen}$.

For our purpose we have to find  the total inclusive cross-section for
production of
exactly $n$ gluons. To this aim we have to integrate
$I_{n}(y_i,k_i)$
both over all $k_i$ and $y_i$.

Integration over momenta was done in ~\cite{bt}.
\beq
I_n(y_i)=\sigma^{(1)}c^n
 e^{\phi(y_i)},
\label{iny}\eeq
where \[\sigma^{(1)}=8\pi\lambda r,\ \
c=
\frac{3\alpha_s}{2\pi^2\Delta},\]
and
\beq\phi(y_i)=
a\Delta^2\Big(2\rho-(n+1)\sigma+n^2\frac{Y}{4}-
\frac{\sigma^2}{Y}\Big)
\label{phi}\eeq
where
\[\sigma=\sum_{k=1}^ny_k,\ \ \rho=\sum_{k=1}^nky_k.\]
So we are left with integration over rapidities.
 to find
\[I_n=\int\prod_{k=1}^ndy_k I_n(y.k).\]
For a particular diagram Fig, 6 the region of integration
$\Omega$ is
\beq\Omega=
\frac{Y}{2}>y_1>y_2>,,,>y_n>-\frac{Y}{2}.\label{omega}\eeq
which generates factor $1/n!$.
However one has $n!$ such diagrams with permutations of $y_i$.
So the result found for the integrated diagram Fig. 3 will carry no
factorial factors.

As stressed in ~\cite{bt} appearance of the exponent $\phi$
lies in fact beyond the LO BFKL model.
Indeed assuming $y_i\sim Y$ and taking into account that
$a\Delta^2\sim \alpha_s^3$ we find that
$\phi\sim\alpha^2(\alpha_s  Y)$, that is   by $\alpha_s^2$ smaller
than terms of the order $(\alpha_sY)^n$ taken into account in the
LO approximation. Therefore in the rigorous treatment one
has to drop the exponential factor in (\ref{iny}) altogether.
However this factor introduces nontrivial $y$- dependence of the
the multiplicity per unit rapidity, apparently
observed in experiment. For this reason we shall consider both cases
with $\phi=0$ and given by (\ref{phi}).

\subsection{Dropping the exponential factor}
In this case
\beq
I_n(y_i)=\sigma^{tot}c^n
\label{iny0}\eeq
and does not depend on rapidities $y_i$.
As for the RG pomeron, factor $c$ has the
 meaning of multiplicity per unit  rapidity.
Remarkably, directly calculated in the
LO BFKL approach, it
does not depend on the QCD coupling constant $\alpha_s$ and  is
just a number (and a rather small one):
\[c=0.0574040...\]
AS mentioned this result describes emission of massless gluons of any
momenta and also does not take into account  energy conservation.
So its relation to the physical multiplicities is not immediate.
The latter,  taking into account the above restrictions and so
with  physical cutoffs, were
studied long ago  in ~\cite{kwiec}.

With (\ref{iny0}) integration over rapidities  is
 trivial and for a particular diagram Fig. 3 give
evidently factor $Y^n/n!$. Multiplying by $n!$  one finds
~\cite{bt}.
\beq
I_n=j^n(Y)\sigma^{tot},\ \ {\rm where}\ \ j(Y)=cY.\label{in0}\eeq
From this as before  we find the Poisson probabilities
for emission from a single pomeron (\ref{poisson})
and continuing the previous derivation  the same
final formulas
for $P(n)$ and its sum for $n>0$ (\ref{pnf}) and (\ref{ratio}).
As before consistency of the model will require $j>>1$. However
this restriction is much more severe for the BFKl pomeron
due to smallness of $c$. Its consequences will be discussed later
in the subsection devoted to numerical calculations.

\subsection{With the nontrivial exponential factor}
Now
\[I_n(y_i)=C^{(n)}e^{\phi(y_i)},\ \ C^{(n)}=\sigma^{(1)}c^n.\]
In this case we meet with certain difficulties in the
integration over  all rapidities.

The first one comes  from the quadratic dependence on
the sum $\sigma=\sum_{k=1}^n y_k$ and the second one from the complicated
integration region $\Omega$.
To make explicit the $Y$-dependence we rescale rapidities
\[y_i=Y\ty_i,\ \sigma=Y\tilde{\sigma},\ \ \rho=Y\tilde{\rho}.\]
In terms of rescaled variables
\beq
\phi(\ty_i)=\gamma(Y)\Big(\frac{n^2}{4}+2\rho-(n+1)\sigma
-\sigma^2\Big),\ \ \gamma(Y)=Ya\Delta^2.
\label{phi1}\eeq

To resolve the first problem we linearize in $\sigma$ introducing
integration over $\sigma$ with the $\delta$-function:
\[\int_{-\infty}^\infty d\sigma\delta(\sigma-\sum_k \ty_k)=
\frac{1}{2\pi}\int_ {-\infty}^\infty dw
\int_{-\infty}^\infty d\sigma e^{iw(\sigma-\sum_k \ty_k)}.\]
We find the integral over $\sigma$, denoting $\gamma_n=(n+1)\gamma$,
\[\int_{-\infty}^\infty d\sigma e^{\sigma(iw-\gamma_n)-\gamma\sigma^2}
=\sqrt\frac{\pi}{\gamma}
\exp\Big(\frac{\gamma_n^2}{4\gamma_n}
-iw\frac{n+1}{2}
-\frac{w^2}{4\gamma}\Big).\]
So we get
\[I_n=C_1^{(n)}
\int_{-\infty}^\infty dw
\exp\Big(-iw\frac{n+1}{2}-\frac{w^2}{4\gamma}\Big)
I^{(y)}_n(w).\]
Here
\[
I^{(y)}_n(w)=Y^n\int_\Omega\prod_{k=1}^ndy_k
e^{\phi_y(y_i,w)},\]
where
\[\phi_y(y_i,w)=2\gamma\rho-iw\sum_{k=1}^ny_k\]
and
\[C_1^{(n)}=C^{(n)} \frac{1}{2\pi}Y^n\sqrt\frac{\pi}{\gamma}
\exp\Big[\gamma\Big((n+1)^2/4+n^2/4\Big)\Big].\]

Now we turn to the complicated domain $\Omega$. To simplify it
we introduce $n+1$ rescaled variables $u_i=y_i-y_{i+1}$,
$i=0,1,..n$,
all positive and
restricted by condition
\beq
\sum_{k=0}^nu_k=1.\label{restru}\eeq

In terms of $u_k$ we find
\beq
\phi_y(u_k,w)=
\sum_{k=0}^n u_k\Big(\gamma k(k+1)-iwk\Big)
-\gamma n(n+1)/2+iwn/2.
\label{phiuk}\eeq
We include the last but one term in the coefficient
and the last one into the integrand in $w$ to obtain
\[I_n=C_2^{(n)}
\int_{-\infty}^\infty dw
\exp\Big(-iw/2-\frac{w^2}{4\gamma}\Big)
I^{(u)}_n(w).\]
Here
\beq
I^{(u)}_n(w)=Y^n\int\prod_{k=0}^ndu_k\delta(\sum_{k=0}^n-1)
e^{\phi_u(u_k,w)},\label{wdelta}\eeq
where
\[\phi_u(u_k,w)=\sum_{k=0}^nu_k(\gamma k(k+1)-iwk)\]
and
\[C_2^{(n)}
=C^{(n)} \frac{1}{2\pi}\sqrt\frac{\pi}{\gamma}e^{\gamma/4}.\]
To integrate over all $u_k$ we impose restriction (\ref{restru})
in the usual manner, introducing the appropriate $\delta$ function
by means of an extra integral
\beq
\frac{1}{2\pi}\int_{-\infty}^\infty d\xi
e^{i\xi (\sum_{k=0}^n u_k-1)}.
\label{inrestr}\eeq
Under this integral one can integrate over positive $u_k$ from zero
to any number $n_{max}\geq 1$. The exponential factor becomes
\[\phi_1(u_k,w,\xi)
=\sum_{k=0}^n u_k\Big(\gamma k(k+1)-iwk+i\xi\Big).
\]
%%%%%%%%%%%%%%%%%%%%%%%%%%%%%%%%%%%%%%%%%%%%%%%%%%%%%%%%%
%%%%%%%%%%%%%%%%%%%%%%%%%%%%%%%%%%%%%%%%%%%%%%%%%%%%%%%%%

We will integrate over $u_k$ up to $u_k=1$.
According to our restriction (\ref{restru}) integration up to
greater values should give the same result.
Integration over all $u_k$   will give
\beq
\prod_{k=0}^n\frac{e^{q_k+i\xi}-1}{q_k+i\xi}.\label{prod}\eeq
where
\[q_k=\gamma k(k+1)-iwk.\]

Now we proceed to integrate over $\xi$. Our integral is
\beq
X_n(w)=\frac{1}{2\pi}(-i)^{n+1}\int_{-\infty}^\infty d\xi e^{-i\xi}
\prod_{k=0}^n\frac{e^{q_k+i\xi}-1}{\xi-iq_k}.\label{xnn}\eeq
The integrand is analytic in $\xi$ but grows exponentially both in the
upper and lower half-planes of complex $\xi$.
Zeros  of the denominator at $\xi=iq_k$ all lie in the upper
$\xi$ half-plane, except for $k=0$. when $q_0=0$. Since  actually
the integrand is analytic at $\xi=0$ we can shift this zero to any
half-plane. It is convenient to shift it also to the upper plane
taking $q_0=+i0$

To manage the integral we add and subtract the constant from
the numerator
and split the integral in two terms
\[X_n(w)=\frac{1}{2\pi}(-i)^{n+1}\int_{-\infty}^\infty d\xi
\Big(F_+(\xi)+F_-(\xi)\Big)\]
where
\[F_+(\xi)=e^{-i\xi}
\prod_{k=0}^n (\xi-iq_k)^{-1}\Big(\prod(e^{q_k+i\xi}-1)-(-1)^{n+1}\Big)\]
and
\[F_-(\xi)=(-1)^{n+1}e^{-i\xi}
\prod_{k=0}^n (\xi-iq_k)^{-1}.\]
Function $F_+(\xi)$ diminishes in the upper half plane but has $n+1$
poles there at $\xi=iq_k$, $k=0,1,2,...n$.
Function $F_-(\xi)$ diminishes
in the lower half-plane and has no poles there. So its integration
over $\xi$ gives zero.
Integration of $F_+(\xi)$ is done by taking residues at its
poles. The product in the numerator of $F_+$ does not give any
contribution and
we get
\beq
X_n(w)
=\sum_{k=0}^ne^{q_k}\prod_{l=0,l\neq k}^n(q_k-q_l)^{-1}.
\label{xn1}\eeq
As expected this result does not depend on the upper limit of
integrations $u_{max}$ over $u_k$ once $u_{max}\geq 1$.
Indeed with $u_{max}$ one will get instead of (\ref{prod})
\beq
\prod_{k=0}^n\frac{e^{u_{max}(q_k+i\xi)}-1}{q_k+i\xi}.\label{prod1}\eeq
This product contains no poles irrespective of the value of
$u_{max}$. After subtraction of  $(-1)^{n_1}$  from the
numerator and multiplied by $\exp(-i\xi)$ it vanishes at large $\xi$
in the upper half-plane if $u_{max}>1$. So repeating our
derivation and taking residues we shall find the same result.

Integration  over $w$ brings us to the final result
\[
I_n=C_2^{(n)}Y^n\int_{-\infty}^\infty dw
\exp\Big(-iw/2-\frac{w^2}{4\gamma}\Big)
X_n(w)\]
\beq
=C_2^{(n)}Y^n\sum_{k=0}^ne^{\gamma k(k+1)}
\int_{-\infty}^\infty dw
\exp\Big(-iw(k+1/2)-\frac{w^2}{4\gamma}\Big)F_k(w)
\label{inexp1}\eeq
where
\[F_k(w)=\prod_{l=0,l\neq k}^n(q_k-q_l)^{-1}\]
and
\[q_k-q_l=\gamma [k(k+1)-l(l+1)]-iw(k-l).\]

Thus  the problem reduces to a complicated one-dimensional
integral in the complex plane. It does not seem possible to find it
analytically.  Its numerical calculations are
complicated by
 violent oscillations of
the integrand. However a shift of the integration variable
\[w\to w-i\gamma(2k+1)\]
allows to overcome this difficulty (see Appendix). One  obtains
\beq
I_n=I_n^{(1)}+I_n^{(2)}\label{inf}\eeq
where
$I_n^{(1)}$ is given by the integral similar to (\ref{inexp1})
but without oscillations
\beq
I_n^{(1)}=
C_2^{(n)}e^{-\gamma/4}\sum_{k=0}^nd_{nk}
\int_{-\infty}^{\infty}
 dw e^{-w^2/4\gamma}{\rm Re}\,\prod_{l=0,l\neq k}^n
 \frac{1}{\gamma(k-l)+iw}\label{inf1}\eeq
and
$I_n^{(2)}$ is an explicit analytical expression
\beq
I_n^{(2)}=
C_2^{(n)}\frac{2\pi(-1)^n}{\gamma^{n-1}}e^{-\gamma/4}
\sum_{k=0}^nd_{nk}
\sum_{m=0}^{k-1}
 (k-m)d_{nm}e^{-\gamma(k-m)^2/4}\label{inf2}\eeq
with the numerical matrix
\[
d_{nk}
=\frac{1}{k!(n-k)!}(-1)^k,\ \ k\leq n\]
The obtained $I_n$ refers to only one particular diagram
Fig. 6. The total $I_n$ is obtained after multiplication by $n!$,
corresponding to permeations of gluons $1,2,...n$.

One observes that while part $I_n^{(1)}$ does not grow with $n$,
part $I_n^{(2)}$ does grow very fast. At large $n$
\[I_n^{(2)}\sim \frac{1}{n!}e^{\gamma n^2/4}.\]
For the commonly assumed value $\alpha_s=0.2$
already at $Y=10$ and $n=20$ part $I_n^{(2)}$ attains values of the order
$10^{+170}$ utterly unacceptable physically.
Of course this is a consequence of the  assumed behavior in the
infrared region of momenta of produced gluons: in spite of all
$I_n$ being stable in the infrared the  bad properties of the model
show themselves in the behavior at large $n$.
This property of $I_n$ prevents construction of probabilities
from $I_n$ using  Eq. (\ref{pvianu}) due to severe divergence at large $k$ of the sum
of $I_k$. Thus unfortunately with the exponential factor present
in (\ref{iny}) one cannot construct reasonable probabilities
although integrated cross-sections $I_n$ can be found by themselves fairly well

To illustrate these points we present in Fig 7. the found $n$-fold
inclusive cross-sections $I_n$ for $n\leq 20$ and $Y\leq 20$
taking somewhat smaller $\alpha_s=0.1$ to make the illustration feasible.
\begin{figure}
\begin{center}
\epsfig{file=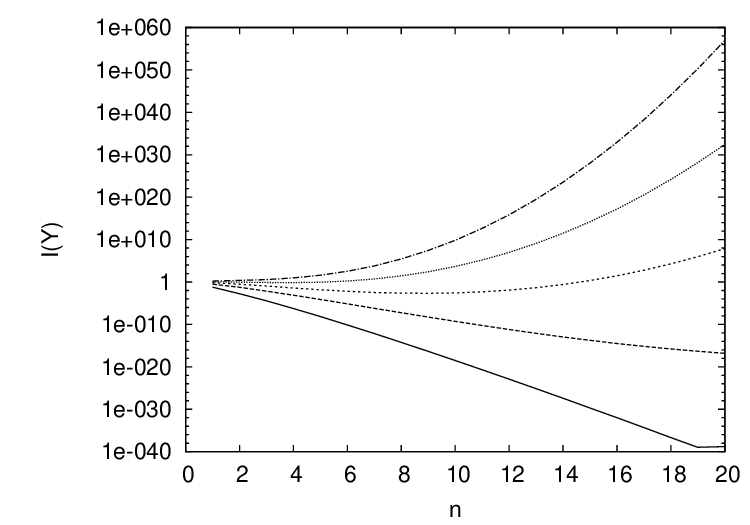, width=12 cm}
\caption{$n$-fold inclusive cross-sections from the BFKL chain as given
by (\ref{iny}) with $\alpha_s=0.1$. Curves from bottom upwards correspond
to $Y=1,5,10,15,20$.}
\end{center}
\label{fig17}
\end{figure}

%%%%%%%%%%%%%%%%%%%%%%%%%%%%%%%%%%%%%%%%%%%%%%%%%%%%%%
%%%%%%%%%%%%%%%%%%%%%%%%%%%%%%%%%%%%%%%%%%%%%%%%%%
%%%%%%%%%%%%%%%%%%%%%%%%%%%%%%%%%%%%%%%%%%%%%%%%%%%%%

\subsection{Numerical calculation}
To have the cross-sections in more or less
agreement with the data we
take  $\alpha_s=0.15$ for  the BFKL pomeron.
For  proton-proton scattering we take $R_1=R_2=0.8$ fm and the
slope $\lambda=2m_\rho^2$. The maximal number of pomerons taken
into account $n_{max}$ was varied from 40 to 3200, as for
the RG pomeron. As for the latter case, for $Y\leq 30$ values
for $P(n)$
at $n\neq 40$ turned out to be already stabilized  at $n_{max}=40$,
whereas for
the entropy only at $n_{max}=800$.

 We find the cross-sections shown in Fig. 8.
\begin{figure}
\begin{center}
\epsfig{file=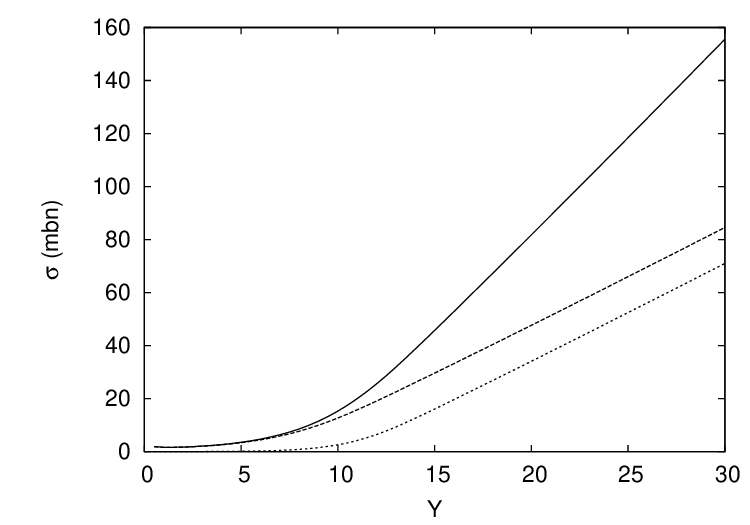, width=12 cm}
\caption{Cross-sections from the BFKL exchanges: total
(upper curve), inelastic (middle curve) and elastic (lowest curve)
 with $\alpha_s=0.15$}
\end{center}
\label{fig8}
\end{figure}

Passing to probabilities $P(n)$ from the start we exclude the
possibility to take into account the nontrivial exponential
factor in (\ref{iny}) and use the emission strength
coefficient $j(Y)=cY$. Then  $P(n)$ and $\sigma^{in}/\sigma^{tot}$
 will be given by the same  formulas (\ref{pnf}) and  (\ref{ratio})
 as for the RG pomeron.
 We retain the natural cutoff $\Delta$ in $c$, so that the value of $c$  is rather small.
 For greater values of $c$ our results will be quite similar to RG pomeron, so that they will not deserve
 a separate study.

Due to the small value of multiplicity per unit rapidity $c$
for the BFKL pomeron  especial importance acquires the consistency
problem, which, we remind, demands
\beq
F\Big(2r(1-e^{-j})\Big)=F(2r).\ \ j=cY.\label{cons2}\eeq
To see the scale of inconsistency
we calculated the  difference of unity of the ratio $R$ of the two
inelastic
cross-sections, from gluon emission and from eikonalization
\[R=\frac{\sigma^{in}_{emission}}{\sigma^{in}_{eikonal}}\]
with the results for $\delta=1-R$ shown in Fig. 9.
\begin{figure}
\begin{center}
\epsfig{file=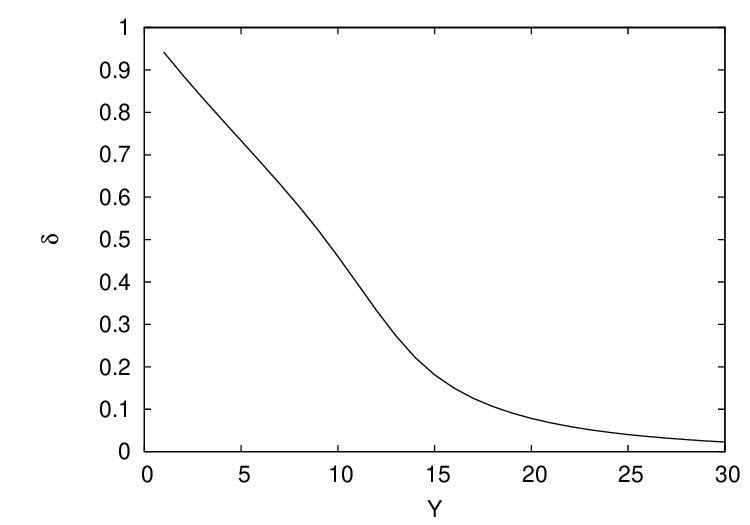, width=12 cm}
\caption{Error $\delta$ in the inelastic cross section at
different $Y$}
\end{center}
\label{fig9}
\end{figure}
As we observe any approximate  consistence of the model
starts for $Y>20$.

In Fig 10 we show our calculated
 probabilities $P(n)$ again multiplied by $\sigma^{tot}$
 for BFKL exchanges at $Y=10,20$ and 30.
\begin{figure}
\begin{center}
\epsfig{file=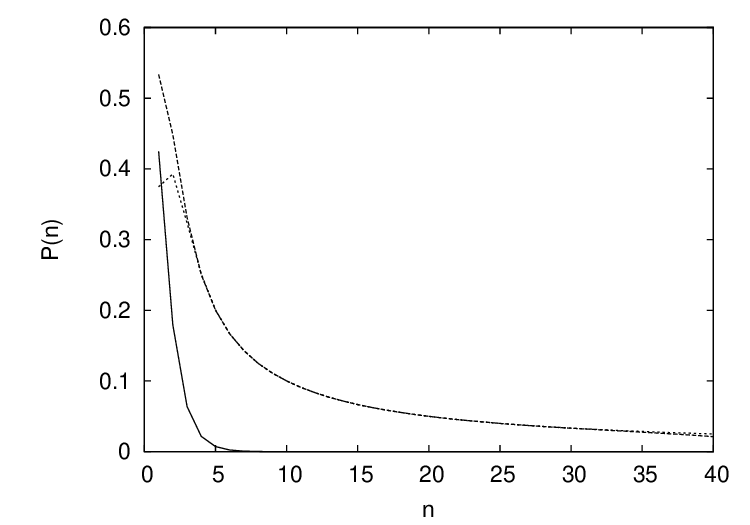, width=12 cm}
\caption{Probabilities $P(n)$ multiplied by $\sigma^{tot}$
for BFKL exchanges at
$Y=10$ (lower curve) $Y=20$ and 30
(merged upper curve for $n>2$).
At $Y=20$ $P(1,2)=0.5335,0. 4485$; at $Y=30$ $P(1,2)=0.3747, 0.3928$.}
\end{center}
\label{fig10}
\end{figure}
Behavior of $P(n)$ for BFKL exchanges significantly changes from
RG ones due to much smaller value of $c$. We do not see any maxima
and the probabilities smoothly (and strongly) diminish with $n$,
Remarkably the probabilities practically do not depend on energy
above $Y=20$ (barring the normalization factor $\sigma^{tot}$)
and so more or less follow the asymptotical trend (\ref{aspn2}).

The found entropy is shown in Fig.11.
Its behavior with rapidity is quite peculiar: it consists of two
distinct parts, the first crudely proportional to $\ln Y$ with a
small slope
and the other  almost exactly proportional; to $\ln Y$ with a
a much larger slope.
\beq S(Y)=0.5\ln Y\theta(14-Y)+4.0\ln Y\theta(y-14).\label{ebfkl}\eeq
Remarkably the transition occurs at rapidities at which the consistency
of the model absent at low $Y$ is practically restored in the vicinity of
$y=14\div 17$. We do not see any deep foundation of this abrupt change
of behavior: it just follows from the property of expression (\ref{pnf})
for the probabilities.
\begin{figure}
\begin{center}
\epsfig{file=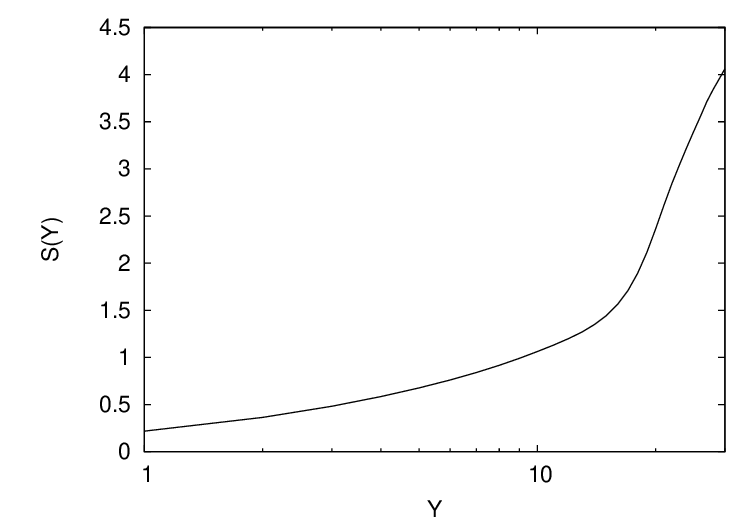, width=12 cm}
\caption{Entropy $S(Y)$ for BFKL exchanges at $Y<30$}
\end{center}
\label{fig11}
\end{figure}

\section{Conclusions}
We studied the possibility to introduce probabilities $P(n)$ in
the pomeron models based on the splitting of the phase space in pieces
referring to $n$ produced particles (gluons). We considered the simplest
model with no interaction between pomeron themselves and so reduced
to multiple pomeron exchanges between participants. This model,
on the one hand, avoids the problem of multiple gluon emission from
pomeron interaction vertices, which prohibits passing to $P(n)$ from the distribution in
cut pomerons. On the other hand, our model is unitary and so allows to see possible
inconsistence found earlier and manifesting itself in the  inadequate inelastic cross section generated by
particle (gluon) production. Our results confirm existence of this inconsistence
and present it in an analytic form. We found that the magnitude of inconsistence diminishes
as a power of energy $s^{-c}$
where $c$ is the parameter characterizing intensity of particle (gluon) production.
So inconsistency  vanishes in the high energy limit and
the model becomes rigorously consistent only in this limit.
The situation at reasonably high energies strongly depends on the value of $c$.
In RG-model with local soft pomerons  $c$ can be chosen more or less
according to the data of the order unity and the inconsistence vanishes very fast. Already at rapidity $Y=10$
it is reduced to 0.01\%. However for the BFKL pomeron $c$ turns out to be nearly 20 times smaller. As  a results
even at $Y=20$ the inconsistence lowers down to only  2\%.

Calculated probabilities $P(n)$ are rather different in these two cases due to different intensity of particle (gluon) production..
For RG pomerons $P(n)$ exhibit clear structure at smaller $n$ with maxima shifted to larger $n$
with energy and only afterwards $P(n)$  go down with $n$. The BFKL exchanges do not show this structure and $P(n)$
smoothly go down with $n$ at all energies (forgetting the inconsistency at  say $Y<20$).
However the calculated entropy as a function of energy has the more or less
similar behavior. In both cases it softly grows with energy as a
logarithm of rapidity. Remarkably   in the
BFKL pomeron case the slope abruptly  grows strongly
at rapidity at which consistency is restored.
In both cases no trace of the asymptotical linear growth in $Y$
found in one-dimensional models is visible.

Further developments in this direction require analysis of multigluon production with interaction between
pomerons taken into account, started with single and double gluon production from the QCD fans
in~\cite{19} and ~\cite{20}. Quite complicated results derived in these papers illustrate that
this formidable task is not to be  easily achieved.

\section{Appendix. Elimination of oscillations}
Presenting
\[I_n=C_2\sum_{k=0}^n\int_{=\infty}^\infty dw e^{\phi_k} F_k(w)=
C_2\sum_{k=0}^n{\rm Re}\,\int_{=\infty}^\infty dw e^{\phi_k} F_k(w),\]
where
\[\phi_k(w)=\gamma k(k+1)-iw(k+1/2)-w^2/4\gamma,\]
we put $w=z-i2\gamma(k+1/2)$.
The exponent $\phi$ becomes
\[\phi_k(z)=\gamma/4+z^2/4\gamma.\]
It does not depend on $k$.

Now we have
\[q_k-q_l=
i(l-k)(z-z_{kl}),\ \
z_{kl}=i\gamma(k-l).\]
So
\[F_k(z)
=\prod_{l=0,l\neq k}^n\frac{1}{(l-k)[\gamma(k-l)+iz]}.\]
We separate from $F_k$ the factor independent of $z$
presenting
\[F_k(z)=d_{nk}G_k(z),\]
where
\[d_{nk}=\prod_{l=0,l\neq k}^n \frac{1}{l-k},\ \ {\rm and}\ \
G_k(z)=\prod_{l=0,l\neq k}^n \frac{1}{\gamma(k-l)+iz}.\]
We find
\beq
I_n=
C_2e^{-\gamma/4}\sum_{k=0}^nd_{nk}
{\rm Re}\,\int_{w_1}^{w_2}
 dw e^{-z^2/4\gamma}G_k(z)\label{eq31}\eeq
where
\[w_1=i2\gamma(k+1/2)-\infty,\ \ w_2=i2\gamma(k+1/2)+\infty.\]
Integration goes along the line parallel to the $x$-axis with
a positive imaginary part $i\gamma(2k+1)$. Of course we have not
excluded oscillations yet, since $z$ has a fixed imaginary part along the
integration line, To eliminate oscillations we have to shift the
integration contour down to the real axis. However  in doing so we
meet poles of function $G(z)$ in the upper half plane of $z$.

These poles occur at $z=z_{km}$
They lie in the upper half-plane for $0\leq m\leq k-1$
but in the lower
one if $m>k$. The pole with a maximal imaginary part occurs for
$m=0$ and is $z_{k0}=i\gamma k$ below the integration line in
(\ref{eq31}).
So shifting the contour to the real axis we have to
add the residues at all $k$  poles in the upper half plane.
We get instead of (\ref{eq31})
\beq
I_n=
C_2e^{-\gamma/4}\sum_{k=0}^nd_{nk}
{\rm Re}\,\Big(\int_{-\infty}^{\infty}
 dw e^{-z^2/4\gamma}G_k(z)+R_k\Big)\label{eq22}\eeq
where $R_k$ is the contribution of poles
\[R_k=-2\pi i(-i)\sum_{m=0}^{k-1}e^{-z_{km}^2/4}
\prod_{l=0,l\neq k,m}^n \frac{1}{\gamma(m-l)}.\]
since at the pole at $z=z_{km}$
\[\gamma(k-l)+iz=\gamma (m-l)\]
Thus
\[R_k=-2\pi i\frac{1}{(\gamma)^{n-1}}\sum_{m=0}^{k-1}e^{-z_{km}^2/4}
\prod_{l=0,l\neq k,m}^n \frac{1}{\gamma(m-l)}.\]
The product can be written as
\[\prod_{l=0,l\neq k,m}^n \frac{1}{m-l}=(-1)^{n-2}(k-m)d_{nm}.\]
We find finally
\[
I_n=
C_2e^{-\gamma/4}\sum_{k=0}^nd_{nk}
\Big(\int_{-\infty}^{\infty}
 dw e^{-z^2/4\gamma}{\rm Re}\,\prod_{l=0,l\neq k}^n
 \frac{1}{\gamma(k-l)+iz}\]\beq
 +2\pi(-1)^n\frac{1}{(\gamma)^{n-1}}\sum_{m=0}^{k-1}
 (k-m)d_{nm}e^{-\gamma(k-m)^2/4}\Big).
\label{eq23}\eeq

Finally the explicit expression for $d_{nk}$.
Dividing the product in factors with $l>k$ and $l<k$
we have
\[d_{nk}=\prod_{l>k}^n\frac{1}{l-k}\prod_{l<k}^n\frac{1}{l-k}
=\frac{1}{(n-k)!}(-1)^k\frac{1}{k!}=\frac{1}{n!}C_n^k(-1)^k.\]

\section{Acknowledgement}
The author is very thankful to the referee for his helpful comment
on the precision of numerical calculations of the entropy.

\section{Declarations}

* Funding --None.

\noindent * Availability of data and material --Not applicable.

\noindent * Competing interests -- Not applicable.

\edo